\documentclass[twocolumn,showpacs,preprintnumbers,amsmath,amssymb]{revtex4}
\usepackage{graphicx}% Include figure files

\begin{document}

\title{Local and Reversible Change of the Reconstruction on Ge(001) Surface between \textit{c}(4$\times$2) and \textit{p}(2$\times$2) by Scanning Tunneling Microscopy}
\author{Yasumasa Takagi, Yoshihide Yoshimoto, Kan Nakatsuji and Fumio Komori}
\address{Institute for Solid State Physics,  University of Tokyo, Kashiwa-shi, Chiba 277-8581, Japan}

\date{\today}

\begin{abstract}
The reconstruction on Ge(001) surface is locally and reversibly changed between \textit{c}(4$\times$2) and \textit{p}(2$\times$2) by controlling the bias voltage of a scanning tunneling microscope (STM) at 80\,K. It is \textit{c}(4$\times$2) with the sample bias voltage $V_b \leq$ - 0.7\,V. This structure can be kept with $V_b \leq$ 0.6\,V.  When $V_b$ is higher than 0.8\,V during the scanning, the structure changes to \textit{p}(2$\times$2). This structure is then maintained with $V_b \geq$ - 0.6\,V.  The observed local change of the reconstruction with hysteresis is ascribed to inelastic scattering during the electron tunneling in the electric field under the STM-tip.
\end{abstract}


\maketitle

Local modification of surface structure in atomic scale has been demonstrated using scanning tunneling microscopes (STM's). Surface adatoms were reversibly manipulated first on a Pt surface at low temperature \cite{Eigler}. On the other hand on semiconductor surfaces, the modifications by temporal voltage pulses, such as the removals of surface atoms \cite{Lyo} and adsorbed hydrogen\cite{Becker,Lyding}, have been performed.  Up to now, however, there has been no report on the reversible modification of local surface reconstructions by using STM.

Such modification can be expected on the ground state \textit{c}(4$\times$2) reconstructed surface of Ge(001) as well as Si(001) because the energy difference between the ground state and the \textit{p}(2$\times$2) structure is estimated to be a few meV/dimer \cite{Needels,Inoue,Yoshimoto}. On the both surfaces, the neighboring two atoms form a buckled dimer, and their reconstructions are characterized by the ordering of the buckled dimers \cite{Chadi}.
Experimentally, fluctuation between \textit{c}(4$\times$2) and \textit{p}(2$\times$2) structures was  induced from  defects on Si(001)\cite{Shigekawa} or from Ag adsorbates on Ge(001)\cite{Naitoh} surfaces.
Recently, the reconstructed structure of a clean Si(001) surface was studied in detail at low temperatures by STM \cite{Hata}. One of the three structures, 2$\times$1, \textit{p}(2$\times$2) and \textit{c}(4$\times$2) is observed, depending both on the sample bias voltage and on the dopant species.
This suggests that the band bending at the subsurface plays an important role to fix the structure as well as the electric field between the STM tip and the surface. It has been argued that the ground state energy of Si(001) surface is partly determined by the electrostatic energy among the electric dipole moments at the dimers \cite{Z}.
In contrast to the extensive studies on Si(001), the ground state of Ge(001) has not yet thoroughly investigated while \textit{c}(4$\times$2) reconstruction was generally observed below 150\,K \cite{Kevan,Kubby}. Up to now,  the transition induced by  the bias voltage change of STM was reported only from \textit{c}(4$\times$2) to \textit{p}(2$\times$2)  on a vicinal Ge(001) surface\cite{Rottger}. 

In the present letter, we show reversible change between \textit{c}(4$\times$2) and \textit{p}(2$\times$2) structures on a clean Ge(001) surface, which can be locally controlled by the sample bias voltage $V_b$ of STM with hysteresis below 80\,K. The \textit{c}(4$\times$2) structure is observed at negative $V_b$ $\leq$ - 0.7\,V while the \textit{p}(2$\times$2) at positive $V_b$ $\geq$ 0.8\,V.  The both structure can be kept even under scanning the surface with  $|V_b|$ $\leq$ 0.6\,V at 80\,K.  These findings are important not only for the fundamental understanding of the stability of the reconstructed surface, but also for the application as a nanoscale memory.

All experiments were performed in an ultrahigh vacuum (UHV) system with a base pressure of below $1\times10^{-8}$\,Pa consisting of a commercial variable temperature scanning tunneling microscope (OMICRON, LT-STM) and a surface preparation chamber \cite{Naitoh}.
Germanium specimens with the size of $8\times{}3\times0.4{}$\,mm$^3$ were cut from a Ge(001) wafer (Sb-dope, 0.35\,$\Omega$$\cdot$cm at room temperature, 0.4 mm thick).
The Ge(001) clean surfaces were obtained by several repetitions of Ar ion
bombardment (1\,keV, 2.5\,$\mu$A$\cdot$cm$^{-2}$, 10\,min) and annealing at 980\,-\,1000\,\,K for 10\,min by passing the DC current directly through the specimen in the UHV preparation chamber. Then, the specimen was transferred to the STM, and the surface was observed at 80\,K or 10\,K with a tungsten tip in a constant current mode. 

Figure 1 shows a series of STM images obtained successively with three values of $V_b$ on the same area of Ge(001) surface at 80\,K. In the case of $V_b$ = - 2.0 \,V the surface superstructure is \textit{c}(4$\times$2) as shown in Fig. 1(a). We can keep this structure when $V_b$  is less than 0.6\,V. Figure 1(b) shows the image with $V_b$ = - 0.2\,V. When we further increase the bias to 1.2\,V,  the structure is \textit{p}(2$\times$2) as shown in Fig. 1(c).
Then, this structure can be maintained with $V_b$ down to - 0.6\,V across the zero-bias as demonstrated in Fig. 1(d) for $V_b$ = - 0.2\,V. The bias dependence of the surface reconstruction at 80\,K is summarized in Fig. 1(e). We can fix either of the two structures at the same bias by changing the bias from $V_b \geq$ 0.8\,V or $V_b \leq$ - 0.7\,V to $|V_b| \leq$ 0.6\,V. It is noted that the both structures can be kept without electric field due to the STM-tip; even after keeping the tip apex at 500 nm away from the area of the interest for several minutes, we observed again the same local structure at the area. The surface structure at 10\,K depends on $V_b$ in the same way as that at 80\,K.
The observed bistability at $|V_b| \leq$ 0.6\,V indicates an energy barrier between the \textit{c}(4$\times$2) and \textit{p}(2$\times$2) structures.
At the structural change, a flip process of the buckled dimers occurs in \textit{c}(4$\times$2) or \textit{p}(2$\times$2) domains. The energy required for such a process in \textit{c}(4$\times$2) domain was theoretically estimated to be 416.7 meV based on a first-principles calculation\cite{Kawai}.
\begin{figure}
\includegraphics[width=7.5cm]{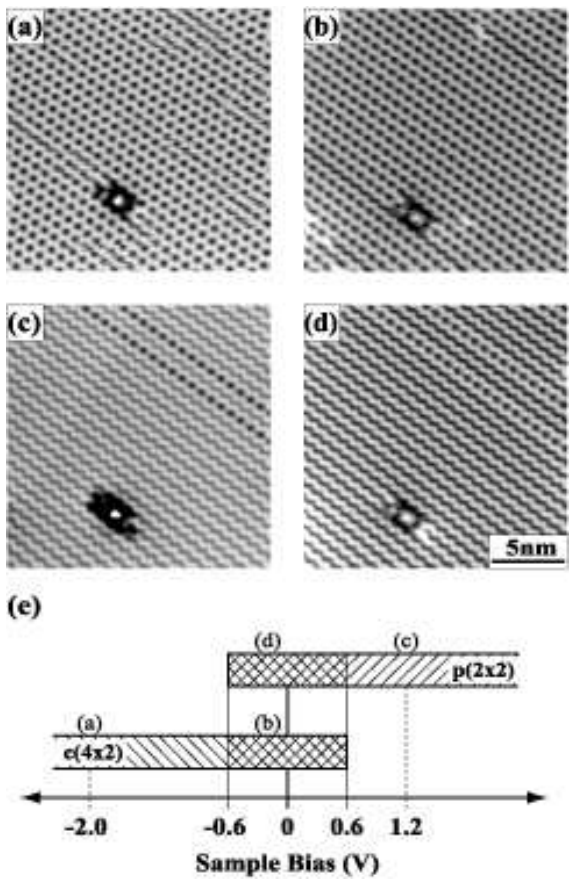}% Here is how to import EPS art
\caption{(a-d) STM images of the same area (18$\times$18\,nm$^2$) on a clean Ge surface at 80\,K with different sample bias voltages; (a) - 2.0\,V, (b) - 0.2\,V changed from a negative bias, (c) 1.2\,V, (d) - 0.2\,V changed from a positive bias. Tunneling current is $I_t$ = 0.43\,nA. (e) A schematic showing the observed superstructures depending on the sample bias voltage and the direction of the voltage change.}
\end{figure}

When we increase $V_b$ in several msec from - 2.5\,V to + 1.2 \,V and continue to scan the surface, the surface gradually transforms from \textit{c}(4$\times$2) to \textit{p}(2$\times$2) through a random series of sudden flip of the  dimers at a single Ge dimer-row as in Figs. 2(a-c). Here, we show selected successive STM images after the $V_b$ change.
Arrows in the figures indicate the sudden changes of the dimer phase during the scanning of the surface from the bottom of the image to the top. When the buckling direction of a dimer changes, all dimers in the same dimer-row change simultaneously. That is, the change of the dimer phase continues to the outside of the scanning area along the dimer row until the dimer row is terminated by steps or any defects.
The transformation of the structure was also observed from \textit{p}(2$\times$2) to \textit{c}(4$\times$2) when we decrease $V_b$, for example,  from + 2.0\,V to - 1.2\,V. In contrast to the change from \textit{c}(4$\times$2) to \textit{p}(2$\times$2), however, this change is limited to a few nm away from the area of the scanning along the dimer-row direction. At the boundaries between the \textit{c}(4$\times$2) and \textit{p}(2$\times$2) areas, there form pairs of the adjacent dimers along the dimer row direction with the same buckling direction.

Figure 2(d) shows a scanning tunneling spectrum spatially averaged on a clean Ge(001) surface at 10\,K. The observed peaks at -1.25\,V and + 1.05\,V are consistent with the previous result at RT \cite{Kubby}.
All the peaks appearing between -1.0 and 1.0\,V are attributed to the dangling bond $\pi$ and $\pi$$^*$ states as in Si(001) surface \cite{Okada}. The positions of the peaks in the tunneling spectra move toward $V_b$ = 0 at 80\,K. This is mainly ascribed to the temperature dependence of the band bending nearby the surface.
\begin{figure}
\includegraphics[width=7.5cm]{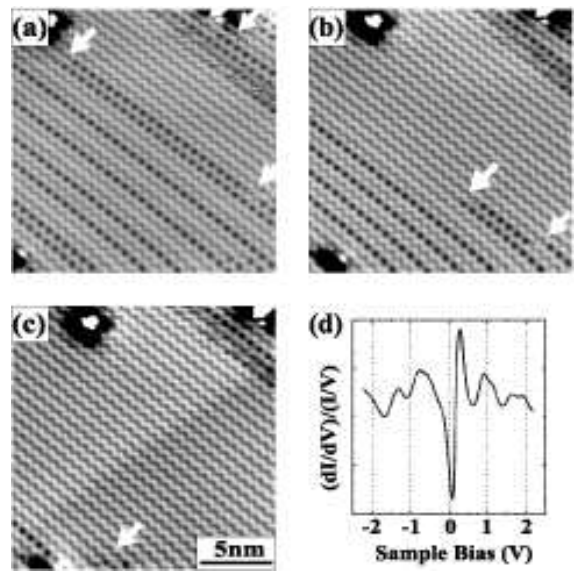}% Here is how to import EPS art
\caption{(a-c) STM images of the same area (20$\times$20\,nm$^2$) of a clean Ge(001) surface. They are selected from the successive images (7 min interval) after the $V_b$ change from - 2.5\,V to 1.2\,V at 80\,K; (a) 1st image (b) 3rd image (c) 7th image. Arrows in the figures indicate the sudden change of the buckling phase of the Ge dimer. (d) A tunneling spectrum of a clean Ge(001) surface at 10\,K.}
\end{figure}

The structures \textit{c}(4$\times$2) and \textit{p}(2$\times$2) are characterized by the phase difference between two adjacent dimer rows. If the dimers are all in phase, the domain structure is \textit{p}(2$\times$2), and out of phase \textit{c}(4$\times$2). Here we define  ``\textit{c}(4$\times$2) length" on an area as the total length of the phase-mismatched adjacent dimers along the dimer-row direction.
We estimated the change rate from \textit{c}(4$\times$2) to \textit{p}(2$\times$2) by measuring the \textit{c}(4$\times$2) length in successive images of the same area as a function of time after a sudden $V_b$ increase at several values of the tunneling current, $I_t$.
The results are shown in Fig. 3 for the $V_b$ change from - 2.0 V to 1.2 V.
It took 6 min to obtain each image of 50 $\times$ 50  nm$^2$. The \textit{c}(4$\times$2) length after the first scanning decreases with increasing $I_t$.  The buckling phases of a few dimer rows on this surface could not be changed by the scanning with $V_b$ = 1.2 V because the phase is firmly fixed by defects on the surface.  Apart from these dimer rows, the characteristic times for the structural change are less than 6 min for $I_t$ =  1.6\,nA and 12 min  for 0.8\,nA.
We confirmed that the tip moves toward the surface just by 0.06 nm with increasing the tunneling current from 0.4\,nA to 2\,nA at $V_b$ = 1.2\,V. This value is much shorter than the tip-surface distance in the tunneling condition. Consequently, the observed structure transformation is little influenced by the changes of the electric field and the atomic force accompanied by the tunneling current change. All these results indicate that the driving force for the change of the buckling phase is mainly attributed to the inelastic electron tunneling.
\begin{figure}
\includegraphics[width=8cm]{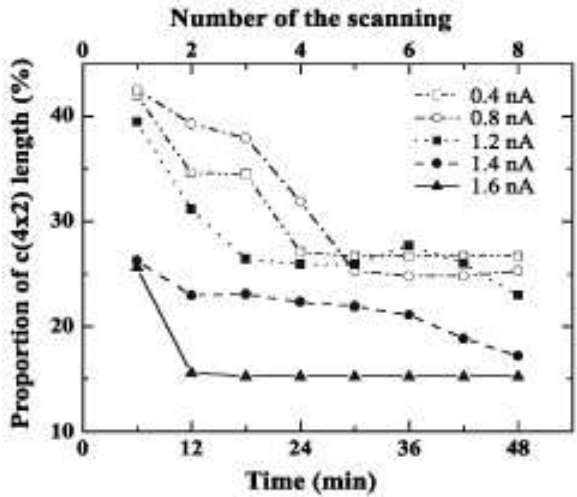}% Here is how to import EPS art
\caption{Proportion of the total length of the phase-mismatched two adjacent dimers indicating the \textit{c}(4$\times$2) superstructure (\textit{c}(4$\times$2) length) in successive eight STM images of the same area after a sudden $V_b$ change from - 2.0\,V to 1.2\,V at five values of $I_t$. The STM images were taken with 6 min interval. During the first scanning, the length decreases rapidly with increasing $I_t$.}
\end{figure}

The phase of the dimer row suddenly changes during the scanning as shown in Fig. 2. The change rate of the structure decreases with decreasing the absolute value of $V_b$ or temperature. We can keep the both superstructures for more than 2 hours during the scanning on the surface, for example, with $I_t$ = 1\,nA and $V_b$ = $\pm$ 0.6\,V at 80\,K. The change rate with $I_t$ = 1\,nA and $V_b$ =  0.8\,V decrease to about 1/3 by decreasing temperature from 80\,K to 10\,K.
According to the tunneling spectrum shown in Fig. 2(d), electrons are effectively injected into or removed from the dangling bond $\pi$ or $\pi$$^*$ states by tunneling when $|V_b| \geq$  0.7\,V. 

Figures 4 demonstrate the successive formations and annihilations of local \textit{p}(2$\times$2) areas by scanning the small area shown as white rectangles at 80\,K. The images were observed with $V_b$ = - 0.2\,V. After taking the image shown in Fig. 4(a), the area A in the figure was scanned with $V_b$ = 0.8\,V to change from \textit{c}(4$\times$2) to \textit{p}(2$\times$2). We reduced $V_b$ as low as possible to avoid any influence to the dimer rows outside of the scanning area. The result is shown in Fig. 4(b).
The change along the dimer row continued to the outside of the scanning area as described above. Then we scanned the area B in Fig. 4(b) to form another \textit{p}(2$\times$2) area. The result is shown in Fig. 4(c). The reverse change from \textit{p}(2$\times$2) to \textit{c}(4$\times$2) was done by scanning the area C in Fig. 4(c) with $V_b$= - 0.7\,V. In this case, we had to scan all of the desired \textit{p}(2$\times$2) area because the change to \textit{c}(4$\times$2) was limited nearby the area of the scanning. The annihilation of \textit{p}(2$\times$2) is shown in Fig. 4(d).
\begin{figure}
\includegraphics[width=6cm]{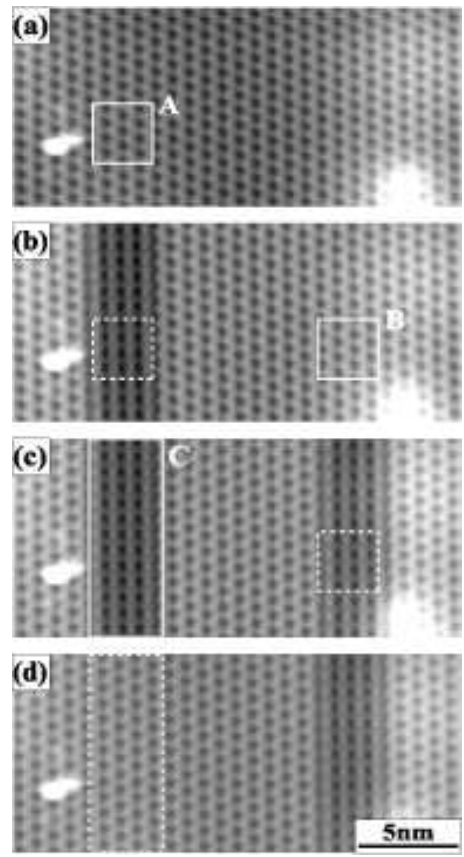}% Here is how to import EPS art
\caption{STM images of the same area (22.5$\times$10\,nm$^2$) of a clean Ge(001) surface at 80\,K, with $V_b$ = - 0.2\,V and $I_t$ = 0.4\,nA. The square areas "A" in (a) and "B" in (b) were scanned with $V_b$ = 0.8\,V to change the local structure from \textit{c}(4$\times$2) to \textit{p}(2$\times$2) after taking each image.  The surface after scanning "A" is shown in (b), and that after scanning "B" is in (c).  The rectangular area "C" in (c) was scanned with $V_b$ = - 0.7\,V to change it from \textit{p}(2$\times$2) to \textit{c}(4$\times$2). The surface after scanning the area "C" is shown in (d).}
\end{figure}

The ground state of Ge(001) surface has been believed to be \textit{c}(4$\times$2) structure.  Both LEED observation\cite{Kevan} and STM study\cite{Kubby} reported \textit{c}(4$\times$2) structure at low temperature consistently with the theoretical studies \cite{Needels,Inoue,Yoshimoto}.
The surface superstructure of Ge(001) is determined by minimizing two competing energies. One is the strain energy at subsurface lattice\cite{Yoshimoto}, and the other is mutual electric-dipole energy among the Ge dimers\cite{Z}.  On this surface, the second-layer atoms under the upper-atom of the dimer are pulled closer while those under the lower-atom are pushed apart to minimize the energy of lattice distortion due to the asymmetric dimer.
In the \textit{c}(4$\times$2) structure, the upper-atom of a dimer is adjacent to the upper-atom in its adjoining dimer row while it is adjacent to the lower-atom in the \textit{p}(2$\times$2) structure. Thus, the strain energy of the \textit{c}(4$\times$2) structure at subsurface is lower than that of the \textit{p}(2$\times$2).
On the other hand, the dipole interaction energy in the \textit{p}(2$\times$2) structure is smaller than that in the \textit{c}(4$\times$2) because the dipoles are oriented to the same direction in \textit{p}(2$\times$2). We note that the upper-atom of the dimer is negatively charged and the lower positively. The \textit{c}(4$\times$2) structure can be the ground state because the energy gain of the lattice-strain dominates the loss of the dipolar energy in this structure. The difference of the energy between \textit{p}(2$\times$2) and \textit{c}(4$\times$2) is as small as 3 meV/dimer\cite{Yoshimoto}. 

When the surface is positively biased under the STM tip, the negatively charged upper-atoms of the dimers are electro-statically pushed toward the subsurface by the tip, and  the lower-atoms of the dimers are pulled toward the surface. Consequently, the dimer buckling becomes smaller. This makes the difference of the subsurface strain energy between \textit{p}(2$\times$2) and \textit{c}(4$\times$2) structures smaller than that of the dipole energy among buckled dimers.
Thus for the positively biased surface, the \textit{p}(2$\times$2) structure is favored, and can be the ground state. On the other hand, under the negative $V_b$, the buckling of the dimers are increased by the electric field, and the \textit{c}(4$\times$2) structure is more favored. This mechanism is qualitatively consistent with the stable structures observed at positive and negative biases by STM.

At present, however, we can quantitatively infer neither the variation of the stability difference between \textit{c}(4$\times$2) and \textit{p}(2$\times$2) structures nor the flipping probability of a dimer by inelastic electron tunneling under the electric field due to the tip-sample bias.
The first-principles calculation of these parameters seems very difficult because of the rather long-scale band bending of Ge subsurface possibly requires extremely large scale calculations, for example. To clarify the transition mechanism in detail,  we should tackle the above two, which are largely dependent on the local electric field and the local current at the dimer.

In case of Si(001) surface, the ground state has been considered to be \textit{c}(4$\times$2) as in Ge(001). Recently, however, both 2$\times$1 and \textit{p}(2$\times$2) structures were observed by STM below 50\,K. Hata \textit{et al}.\cite{Hata} claimed that the observed 2$\times$1 structure at 10\,K is attributed to the STM-tip induced flip-flop of the Si dimers. In this case, \textit{p}(2$\times$2) structure appears only in n-type Si, and \textit{c}(4$\times$2) only in p-type Si.
Moreover, the latter was observed only in a very small range of $V_b$. These are quite in contrast to the present results at Ge(001) surface. We have never observed 2$\times$1 structure or symmetric dimers on Ge(001) surface below 80\,K except at the dimer rows adjacent to defects on the surface. The both \textit{c}(4$\times$2) and \textit{p}(2$\times$2) structures are observed in a wide $V_b$ range, and are stable during the STM observations with $V_b$ between - 0.6\,V and 0.6\,V.  The difference from Si(001) surface is ascribed to the large energy barrier between \textit{c}(4$\times$2) and \textit{p}(2$\times$2) structures in Ge(001) \cite{Inoue,Yoshimoto}.

In conclusion, we have demonstrated the local modification of the Ge(001) reconstruction  between \textit{c}(4$\times$2) and \textit{p}(2$\times$2) reversibly by $V_b$ change using STM below 80\,K. The \textit{p}(2$\times$2) structure can be the ground state under an electric field, and inelastic tunneling is attributed to the driving force to the local transformation of the structure. The two different structures coexist with $|V_b| \leq$ 0.6\,V, which is applicable to a nanoscale memory.

\end{document}